
\input harvmac

\def\const {{\rm const}}
\def \s {\sigma}
\def\t {\theta}

\def \tt {\tilde \theta}

\def \p {\phi}
\def \ha {\half}
\def \ov {\over}
\def \fourth  {{1\ov 4} }
\def \four {\textstyle {1\ov 4}}
\def \a {\alpha}
\def \lr { \lref}
\def\ep{\epsilon}

\def\vp {\varphi}
\def \bd {\bar \del}
\def \r {\rho}
\def\const {{\rm const}}\def\bd {\bar \del} \def\m{\mu}\def\n {\nu}\def\l
{\lambda}

\def \tip { \tilde \vp}
\def \y { \tilde \y}

\def \L {\Lambda}
\def\y {{ \tilde y}}

 \def \sm {$\s$-model\ }

\def \lr { \lref}

\gdef \jnl#1, #2, #3, 1#4#5#6{ { #1~}{ #2} (1#4#5#6) #3}

\def\np {  Nucl. Phys. }
\def \pl { Phys. Lett. }

\def \prl { Phys. Rev. Lett. }
\def \pr  { Phys. Rev. }
\def \prd  { Phys. Rev. }

\baselineskip8pt
\Title{\vbox
{\baselineskip 6pt{\hbox{ }}{\hbox
{Imperial/TP/94-95/7 }}{\hbox{hep-th/9411198}} {\hbox{revised}}} }
{\vbox{\centerline { Melvin solution in string theory }
\centerline {  }
}}

\vskip -20 true pt

\centerline{   A.A. Tseytlin\footnote{$^{\star}$}{\baselineskip8pt
e-mail address: tseytlin@ic.ac.uk}\footnote{$^{\dagger}$}{\baselineskip8pt
On leave  from Lebedev  Physics
Institute, Moscow, Russia.} }

\smallskip\smallskip
\centerline {\it  Theoretical Physics Group, Blackett Laboratory}
\smallskip

\centerline {\it  Imperial College,  London SW7 2BZ, U.K. }
\bigskip
\centerline {\bf Abstract}
\medskip
\baselineskip8pt
\noindent
We identify   string theory
counterpart  of the dilatonic Melvin $D=4$ background
describing a   ``magnetic flux tube" in low-energy field theory limit.
The corresponding  $D=5$ bosonic string model containing extra compact
Kaluza-Klein
dimension  is a direct product of
the $D=2$  Minkowski space and  a $D=3$  conformal $\s$-model. The latter is
a singular  limit of the $[SL(2,R)\times R] /R $ gauged WZW  theory.
This implies, in particular,  that the dilatonic Melvin background  is an exact
string solution to all orders in $\a'$.  Moreover,
the $D=3$ model  is formally  related by an abelian duality to a  flat space
of  non-trivial topology.
    The  conformal field theory for the Melvin solution
is   exactly solvable
 (and for special values of magnetic field parameter  is equivalent to CFT for
 a  $Z_N$ orbifold of  2-plane times  a circle)
and should  exhibit   tachyonic instabilities.

\Date {November 1994}

\noblackbox
\baselineskip 16pt plus 2pt minus 2pt
\lr \mel { M.A. Melvin, \pl 8 (1964) 65; W.B. Bonner, Proc. Phys. Soc. London
A67 (1954) 225. }

\lr\mono{T. Banks, M. Dine, H. Dijkstra and W. Fischler, \pl B212 (1988) 45.}
\lr\rb {I. Robinson, Bull. Acad. Polon. Sci. 7 (1959) 351;
B. Bertotti, \pr 116 (1959) 1331. }

\lr\gaun { H.F. Dowker, J.P. Gauntlett, D.A. Kastor and J. Traschen,
\prd D49 (1994) 2909;
 H.F. Dowker, J.P. Gauntlett, S.B. Giddings and G.T. Horowitz, \prd D50 (1994)
2662; S.W. Hawking, G.T. Horowitz and  S.F. Ross, ``Entropy, area and black
hole pairs", NI-94-012, DAMTP/R 94-26, UCSBTH-94-25, gr-qc/9409013. }

 \lr \napwi {C. Nappi and E. Witten, \prl 71 (1993) 3751.}
\lr\tset{A.A. Tseytlin, \pl B317 (1993) 559.}
\lr \ruts { J. Russo and A.A. Tseytlin, ``Constant magnetic field in closed
string theory: an exactly solvable model", CERN-TH.7494/94,
Imperial/TP/94-95/3, hep-th/9411099. }

\lr \givki{A. Giveon and E. Kiritsis,  \np B411 (1994) 487.}
\lr\sft{  K. Sfetsos, \pl B324 (1994) 335.}
\lr\sfts{K. Sfetsos and  A.A. Tseytlin, \np B427 (1994) 325.}
\lr\kum{A. Kumar, \pl B293 (1992) 49; S. Hassan and A. Sen,
\np B405 (1993) 143; E. Kiritsis, \np B405 (1993) 109. }

\lr \tsee{A.A. Tseytlin, ``Exact string solutions and duality", to appear in:
{\it Proceedings of the 2nd Journ\' ee Cosmologie}, ed. H. de Vega  and N. S\'
anchez (World
Scientific), hep-th/9407099. }
\lr \hot { G.T. Horowitz and A.A. Tseytlin, \pr D50 (1994) 5204. }

\lr \alv {
A. Giveon and M. Ro\v{c}ek, Nucl. Phys. B421 (1994) 173;
E. \'Alvarez, L. \'Alvarez-Gaum\'e, J.L.F.
Barb\'on and Y. Lozano, \np B415 (1994) 71;
A. Giveon, M. Porrati and E. Rabinovici, Phys. Rep. 244
(1994) 77.}

\lr\ghs { D. Garfinkle, G.T. Horowitz and A. Strominger,  \pr  D43 (1991) 3140;
1991; {D45} (1992) 3888(E).}

\lr\busch{T.H. Buscher, Phys. Lett. B194 (1987) 51; B201 (1988) 466.}

\lr \rocver{M.  Ro\v cek  and E. Verlinde, \np B373 (1992) 630.  }

\lr \dab {A. Dabholkar, ``Strings on a cone and black hole entropy",
HUTP-94-A019,  hep-th/9408098; ``Quantum corrections to black hole entropy in
string theory",    hep-th/9409158.}
\lr \stro{D.A. Lowe and A.  Strominger, ``Strings near a Rindler or black hole
horizon", UCSBTH-94-42, hep-th/9410215.}

\lr\hrt {G.T. Horowitz and A.A. Tseytlin. \pr D50 (1994) 5204. }
\lr\bag{ J.A. Bagger, C.G.  Callan and J.A.  Harvey, \np B278 (1986) 550. }

\lr\ant{I. Antoniadis, C. Bachas and A. Sagnotti, \pl B235 (1990) 255.}
\lr\bak{C. Bachas and E. Kiritsis, \pl B325 (1994) 103. }

\lr\rutsn{J. Russo and A.A. Tseytlin, to appear }

\lr \kltspl { C. Klim\v c\'\i k and A.A. Tseytlin, \pl B323 (1994) 305.}

\lr \gibma { G.W.  Gibbons and  K. Maeda, \np B298 (1988) 741.}
\lr \gibb{
G.W.  Gibbons, in: {\it Fields and Geometry}, Proceedings of the 22-nd Karpacz
Winter School of Theoretical Physics, ed. A. Jadczyk (World Scientific,
Singapore,  1986).}

\lr \gps {S.  Giddings, J. Polchinski and A. Strominger,   \pr D48 (1993)
 5784. }

\lr\horts{ G.T. Horowitz and   A.A. Tseytlin, ``A new class of exact solutions
in string theory", Imperial/TP/93-94/54, UCSBTH-94-31, hep-th/9409021.}

\lr \hoho {J. Horne and G.T. Horowitz, \np B368 (1992) 444. }
\lr \sftse { K. Sfetsos and A.A.  Tseytlin,  \pr D49 (1994) 2933.}

\lr\khu{R. Khuri, \pl B259 (1991) 261; \np B387 (1992) 315.}

\lr \nels {W. Nelson, \pr D49 (1994) 5302.}
\lr \kalor { R. Kallosh and T. Ort\'\i n, ``Exact $SU(2)\times U(1)$ stringy
black holes", SU-ITP-94-27, hep-th/9409060. }

\lr \busc {T.H. Buscher, \pl  B194 (1987) 59; \pl B201 (1988) 466;
 M. Ro\v cek and E. Verlinde, \np B373 (1992) 630. }

\lr \los { D.A. Lowe and A. Strominger, \prl 73 (1994) 1468.}

\lr \kk {E. Kiritsis and  C. Kounnas,  \pl B320 (1994) 361.}
\lr\sus{ L. Susskind, ``Some speculations about black hole entropy in
string theory", RU-93-44 (1993), hep-th/9309135. }

\vfill\eject


1.\  There are two  simple  analogues  of a uniform   Maxwell
  magnetic  field background
  in  the Einstein-Maxwell
theory:  one is the Robinson-Bertotti solution  \rb, i.e. (AdS)$_2 \times S^2$
with  covariantly constant  magnetic  field   $F=B_0\sin \t d\t \wedge d\vp$
(``monopole")  on $S^2$,
and another is  the static cylindrically symmetric
 Melvin magnetic universe (or ``flux tube")  \mel.
The latter solution  has $R^4$  topology  and  may be used to describe
  a background magnetic field.
 Several interesting  features  of the Melvin  solution in the context of
Kaluza-Klein (super)gravity (e.g. instability against monopole  or  magnetic
black hole pair creation)  were emphasized  in \gibb\ (see also  {\gaun}).

Both solutions have  direct  generalisations to
low-energy string theory (heterotic string or $D=5$  bosonic string
compactified to $D=4$). In addition to the
 metric and  the
vector field
the string effective action
 includes  also the dilaton $\p$
and the antisymmetric tensor $B_{\m\n}$.
The  analogue of the Melvin solution  \gibma\ has $B_{\m\n}=0$ but  a
non-constant
dilaton. As was noted   recently \refs{\horts,\ruts},
 in the context of   string theory
there is   also {\it another}
  natural uniform magnetic field  solution.
For   $D=4$ its metric ($ds^2= - (dt + A_i dx^i)^2 + dx_idx^i + dz^2, \ \ A_i=
-\ha F_{ij}
x^j, \ \  i=1, 2$)
is that of a product of a real line $R$ and the  Heisenberg group space
$H_3$, the dilaton is constant but  the antisymmetric tensor field strength
is non-trivial and is equal to the  (covariantly) constant magnetic field
$H_{0ij}= F_{ij}=\const $.  This background is an exact string solution \horts\
and the corresponding conformal string model is easily   solvable \ruts.
The  Robinson-Bertotti solution also has an exact string
 counterpart \los\
which is a product of the two conformal theories:
``(AdS)$_2$"   ($SL(2,R)/Z$ WZW)   and
``monopole"   ($SU(2)/Z_m$ WZW)  \gps\  ones.\foot{Kaluza-Klein monopoles in
string theory were considered in \mono.
To use $SU(1,1)$
WZW model to construct (electro)magnetic string  backgrounds by dimensional
reduction was suggested in \ant.
An exact  $D=3$ monopole-type \khu\
magnetic background based on $SU(2)$ WZW model tensored with linear dilaton
 was  discussed   in \bak.}

This raises   the question  we address below: which is  an exact string
counterpart
of the leading-order  dilatonic Melvin background?
 Starting with the $D=5$ bosonic string effective action and
 assuming that one spatial dimension $y=x^5$ is compactified on a circle,  one
finds the following dimensionally reduced  $D=4$  action
\eqn\acttp{  S_4 = \int d^4 x \sqrt {\hat G }\  e^{-2\phi + \s}    \ \big[
  \   \hat R \ + 4 (\del_\m \phi )^2 - 4 \del_\m \phi \del^\m \s }
$$  - {1\ov 12} (\hat H_{\m\n\l})^2\  - \fourth e^{2\s} ({  F}_{\m\n}
({\cal A}))^2
-\fourth  e^{- 2\s} (F _{\m\n} ({\cal B}))^2
  + O(\a')   \big]  \  , $$
where
\eqn\fgfg{\hat G_{\m\n} \equiv  G_{\m\n} - G_{55}{\cal A}_\m {\cal A}_\n
\ , \ \ \ G_{55}\equiv  e^{2\s}\ , \ \ \ \
\hat H_{\l\m\n} = 3\del_{[\l} B_{\m\n]} - 3 {\cal A}_{[\l} F_{\m\n]}
({\cal B})
\ ,  }
  $$   F_{\m\n} ({\cal A}) = 2\del_{[\m}
{\cal A} _{\n]}  \ ,   \ \ F _{\m\n}({\cal B}) = 2 \del_{[\m} {\cal B}
_{\n]}  \  ,
\ \ \  {\cal A} _\m\equiv   G^{55}  G_{\m 5}\ ,  \ \ {\cal B} _\m \equiv
B_{\m 5}\  .  $$
For both Melvin solution \gibma\ and the solution of \horts\
$\s=0 $ so that the two vector fields  ${\cal A} _{\n}$
and ${\cal B} _{\n}$  are equal (up to sign). Then \acttp\ reduces to
\eqn\app{  S_4 = \int d^4 x \sqrt {\hat G }\  e^{-2\phi}    \ \big[
  \   \hat R \ + 4 (\del_\m \phi )^2   - {1\ov 12} ( H_{\m\n\l})^2\  - {1\ov 2}
({ F}_{\m\n}
( { A}))^2 + O(\a')   \big]  \  , }
where we have set  $A_\m =  {\cal A}_\m= -{\cal B}_\m$
 (the Chern-Simons term vanishes in the axially symmetric  case  we consider
so we omit the hat on $H$).
  In this string frame the dilatonic Melvin solution has  a  simple form
 \gibma\ (here $D=4, \ a=1$)\foot{In the Einstein frame the  magnetic pressure
of the flux tube  is  balanced by the gravitational and dilatonic attraction.
In the string frame the dilatonic contribution has the same sign as the
magnetic field one.}
\eqn\melv{ ds^2_4= \hat G_{\m\n}dx^\m dx^\n = - dt^2  + dz^2 + d\r^2 +  \L^{-2}
 (\r)  \r^2 d\vp^2\ , }
\eqn\me{ A= A_\m dx^\m  =  b \L\inv (\r)   \r^2 d\vp\ , \ \ \   F^2_{\m\n} =
8b^2 \L^{-2} (\r) ,  }
\eqn\mme{  \p=\p_0 - \ha \ln \L (\r)  \ , \ \ \  \ \ \
    \L (\r) \equiv 1 +   b^2 \r^2 \ , \ \ \ b=\const \ .   }
The parameter $b$ determines the strength of the magnetic field.
 The curvature,  magnetic field and the  effective
string coupling  $ e^{\p}$
have maxima at $\r=0$ and decrease to zero at infinity.\foot{Since the
radius of the circular $\vp$ dimension approaches zero at infinity
this solution was also interpreted  in \gibma\ as a ``Minkowski membrane"
with   the $(\r,\vp)$-space (a ``bottle" with an infinite narrowing  neck)
 playing the role of a  non-compact internal 2-space. }
{}From the point of view of the low-energy field theory one could formally  use
also the
gauge potential
\eqn\mee{ A'=  A - b\inv d\vp= - b\inv \L\inv (\r) d\vp
  \ ,   }
which  has the same field strength but is singular on the $\r=0$ axis  (the
gauge transformation
is singular at $\r=0$). The string models corresponding to  $A$ and $A'$,  in
general,  will
{\it not}  be equivalent (see below).

For comparison,
 the constant   magnetic field
solution of \app\  found in \horts\ has the following explicit form
\eqn\hts{ ds^2_4 = - (  dt + b \r^2 d\vp)^2 +  dz^2 + d\r^2 + \r^2 d\vp^2\ , }
\eqn\meh{ A =  b  \r^2 d\vp\   , \ \ \ \ \  \ \ \  F^2_{\m\n} =  8b^2  \ ,  }
\eqn\mmeh{   B= \ha B_{\m\n} dx^\m \wedge dx^\n =  b  \r^2 d\vp\wedge dt\ ,
\ \ \  H= F \wedge dt \ , \ \ \ \p=\const   \ .  }
While the metric  \melv\ is static and  is a product of the $D=2$   Minkowski
space and a curved $D=2$ part, the metric \hts\ is stationary
with a non-trivial part being 3-dimensional.

\bigskip
2.  \ As in the case of \hts,\meh,\mmeh\  discussed  in \refs{\horts,\ruts}
we  can now write down the  $D=5$ string $\s$-model   which corresponds
to the dilatonic Melvin solution \melv\ (considered
as  a $D=4$ ``image"  of a $D=5$ bosonic string  background). Using the
correspondence between
\acttp\ and \app\  we get
 \eqn\lagr{ I_5={1\over \pi\alpha '}\int d^2 \s\big[  (\hat G_{\m\n} +B_{\m\n})
\del x^\m \bd x^\n + \  e^{2\s} [\del y+ {\cal A}_\m (x) \del x^\m][\bd y+
{\cal A}_\n (x)
\bd x^\n] } $$
 + \  {\cal B}_{\m   } (x) (\del x^\m \bd y- \bd x^\m \del y )  + {\cal R} \p
(x) \big]\ , \ \ \ \ {\cal R}\equiv  \fourth  \a'R^{(2)} \ ,  $$
i.e.,
\eqn\nnnn{ I_5=  {1\over \pi\alpha '}\int d^2 \s\big(  -  \del t \bd t
 +   \del z \bd z +   \del \r \bd \r  +  \L^{-2}  (\r)  \r^2 \del \vp \bd \vp
}
$$  + \ [\del y+  b \L\inv (\r)   \r^2 \del \vp ][\bd y+  b \L\inv (\r)   \r^2
\bd \vp ] $$  $$
 + \   b \L\inv (\r)   \r^2 (\bd \vp \del y- \del \vp \bd y )  + {\cal R}
[\p_0 - \ha  \ln \L (\r)]\  \big)\ .  $$
Here $\vp$ has the   period $2\pi$ and  the ``Kaluza-Klein" coordinate $y$ --
the period $2\pi R$. The action \nnnn\ can be represented as the sum of
$(t,z)$   Minkowski part and a non-trivial 3-dimensional $(\r,\vp,y)$
theory
\eqn\nnn{ I_5=  {1\over \pi\alpha'}\int d^2 \s\big(  -  \del t \bd t
 +   \del z \bd z  +   \del \r \bd\r +  \L^{-1}
(\r)  \r^2 \del \vp \bd \vp   +  \del y \bd y  }   $$
 + \  2  b \L\inv (\r)   \r^2 \bd \vp \del  y \  +\  {\cal R}[\p_0- \ha   \ln
\L (\r)  ]\ \big)\equiv  {1\over \pi\alpha'}\int d^2 \s\big(  -  \del t \bd t
 +   \del z \bd z  ) + I_3 \ . $$
 If one uses   $ A'$ in \mee\ the action found from \lagr\  is
\eqn\nn{ I'_5=  {1\over \pi\alpha'}\int d^2 \s\big(  -  \del t \bd t
 +   \del z \bd z  +   \del \r \bd\r +  b^{-2}
\L^{-1}  (\r)  \del \vp \bd \vp    +  \del y\bd y  }
$$ - \  2  b\inv  \L\inv (\r)    \bd \vp \del  y  + {\cal R}[\p_0 - \ha   \ln
\L (\r)] \big)\ .   $$
The actions \nnn\  and \nn\ are  formally related by the ``gauge
transformation"
\eqn\tra{ y\to  y'=  y - b\inv \vp  \ , }
which is well-defined only if the periods of $y$ and $b\inv \vp$ are the same.
Since \nnn\ is  non-singular
this  suggests  that \nn\ defines a regular  string model only if
$R=m R', \ b\inv = nR'\ $ ($m, n$ are integers and $2\pi R'$ is a period of
$y'$),
i.e. when  $Rb= m/n$ (see also below).
The string models  \nnn\ and \nn\ are  thus
 equivalent  ($R=R'$) only
if $Rb=1/n$. It is the non-singular model \nnn\ that is  the string analogue of
the Melvin solution.

Eq. \nnn\  may be compared with the model
\eqn\xxx{ {\bar I}_5=  {1\over
\pi\alpha '}\int d^2 \s\big[  -  \del t \bd t
 +   \del z \bd z +   \del \r \bd \r  +    \r^2 \del \vp \bd \vp  }
$$+ \ \del y \bd y + 2b\r^2  \bd \vp  (\del y -  \del t) \  + \ {\cal R}\p_0
\big]\ ,   $$
representing  the solution \hts,\meh,\mmeh\ \refs{\horts,\ruts}.
Near $\r=0$ (where the magnetic field is approximately  constant)
\nnn\ reduces  to \xxx\ up to the ``rotation"
 term  $-2b\r^2  \bd \vp  \del t$.

\bigskip
3. \  Let us now show that  the non-trivial $(\r,\vp,y)$ part $I_3$ of \nnn\
is related by  a
special limit to  the $[SL(2,R)\times R ]/R $ gauged WZW  (or ``charged black
string") model \hoho.
A (locally) equivalent  $D=3$  model appeared in \sft\
(it was obtained by gauging a $U(1)$ subgroup
 of the non-semisimple $E^c_2$ WZW model of \napwi)  where
its formal relation to $[SL(2,R)\times R ]/R $ gauged WZW  model
 was already observed.\foot{I am grateful to K. Sfetsos for pointing this out
to me while the present paper was in preparation.}
 In  the Euler angle
parametrisation of $SL(2,R)$,
$ \ g= {\rm e}^{{i\over 2 } \t_L \s_2 } {\rm e}^{{1\ov 2}
{\tilde r}\s_1} {\rm e}^{{i\over 2 } \t_R \s_2  }, \ \t_L= \t + \tt, \ \t_R=
\tt-\t, $ the action of the $[SL(2,R)\times R ]/R $ gauged WZW  model  is given
by
(see e.g. \sftse)
\eqn\wzw{ I_{gwzw}(r, \t,\tt)
 = {k \over  \pi } \int d^2 \s \big(\
\four \del r\bd r\  +  (1 + q)  {[1-2(1+q)X\inv (r)] } \del \t\bd \t}
 $$ -q {[1-2q X\inv (r)] }  \del \tt \bd \tt
-  2q(q+1)X\inv (r) ( \del \t \bd \tt  -  \del \tt \bd \t )
\  + \   {\cal R} [\p_0'- \ha   \ln X(r)]  \big)  , $$ $$ \ \ \ \ X(r) \equiv
\cosh r + 1+2q
\ .     $$
 The free
parameter $q$  determines the embedding of $R$ into $SL(2,R)\times R$.
It is easy to see that $I_3$ in \nnn\   can be formally
  obtained  from \wzw\ in  the
following special limit
(in this  limit   $X(r) \to  2b^{-2} \ep^2 \L(\r)$)\foot{The mass and axionic
charge of the  black string \hoho\ are related to $k$ and $q$
by $M= q/\sqrt k  , \ Q^2= {{q(1+q)}/k}$  and thus vanish  in  this limit
 but the simultaneous  singular rescalings of the coordinates
give  rise to a non-trivial  model (see also \sft).}
\eqn\limi{ k=(\a'\ep^2 )^{-1} \to \infty  \  , \
\  \  \  q= -1 + b^{-2} \ep^2\ \to -1 \  , \ \ \ e^{\p_0'}=2b^{-2} \ep^2
e^{\p_0}\ , \ \ \  \ep\to 0 \ ,  }
\eqn\lii{ r= 2\ep\r\ , \ \ \  \ \ \  \tt= \ep [-(q+1)/q]^{1/2} y\  \to\
 b^{-1} \ep^2 y\  , }
$$ \t= \vp  + {q\ov q+1} \tt = \vp + \ep [-q/(q+1)]^{1/2}\  \to  \ \vp +  by \
{}.
  $$
This  limit does  $not$, however,   respect the {\it global}  structure of the
two models:
since $\t, \tt$ and $\vp$ should  have periods $2\pi$,
the radius   $ R$ of $y$  should satisfy  $ b^{-1} \ep^2 R= m$ and
$bR=n$ which is not possible for integer $m$ and $n$.
Still,  it  preserves the local  operator  relations
of the coset CFT (like the expression for  the stress tensor in terms of
currents and OPE's), i.e.   gives rise to  analogous  relations for the
theory \nnn.  The  CFT corresponding to  \nnn\ is, in fact,  much simpler
than (any regular limit of)
 $[SL(2,R)\times R]/R $  coset model.  For example,
its central charge  has   the free-theory value (since $c_{gwzw} = 3k/(k-2)$
does not depend on $q$ its limit is  simply  the $k\to \infty$ one).
The existence of this formal limiting relation  is  also sufficient
to argue that   \nnn\ is  a conformal $\s$-model.
The fact that in  the limit \limi\   $k\to \infty$
does not by itself
 imply that all higher  loop corrections to $\beta$-functions
 should  automatically vanish  since they may
depend on $q$ in a general scheme (see \sftse). However, for any $q$
there exists a scheme in which the leading-order background corresponding to
\wzw\
is exact to all orders.\foot{  For the
$[SL(2,R)\times R]/R$ model
 this was explicitly checked at  the two-loop order in \sftse\ and
was also  suggested  in the context of
 conformal perturbation theory
 in  \givki.
Similar statement holds  for  other   gauged WZW backgrounds
as was demonstrated to all orders
on the example of $SL(2,R)/R$ in
\tset\ and argued to be true in general in \refs{\hot,\tsee}.}
We thus conclude that (in such scheme)
the dilatonic Melvin solution
\melv,\me,\mme\ (embedded in bosonic string theory according to \nnnn)
is an {\it exact
string solution}.

\bigskip
3. \ A simple structure    of the Melvin model  (suggested  by its $k\to
\infty, q\to -1$ relation to  the
$[SL(2,R)\times R]/R$   coset model)
is further revealed
 by  making  abelian  duality transformations \busch.
This  helps  to construct explicitly
the corresponding conformal field theory
 since abelian duality in the direction of a compact
isometry gives a \sm representing an  equivalent CFT \refs{\rocver,\alv}.
We shall see that duality in the  angular coordinate of  the plane
leads to non-trivial
 models when combined with shifts in the compact Kaluza-Klein dimension.

The action  \nnn\ has two non-trivial symmetries
under  constant shifts of $y$ and $\vp$.  Like \xxx\ the  model \nnn\ is
``self-dual" with respect to  duality in  the $y$-direction
(the dual action has the same form with $y$ replaced by the  dual coordinate
$\y$ with period $2\pi \tilde R, \  \tilde R= \a'/R$).
The duality transformation in the $\vp$ direction can be performed
by gauging the symmetry  $\vp\to \vp + a$ of \nnn\  \refs{\rocver,\alv}
\eqn\nng{ I_3=  {1\over \pi\alpha'}\int d^2 \s\big(   \del \r \bd\r +
\L^{-1}  (\r) \r^2  (\del \vp  + V) (\bd \vp  + \bar V)
   +  \del y \bd y  }
 $$ +   \ 2 b  \L\inv (\r)   \r^2  (\bd \vp  + \bar V) \del  y
+ \a'(\bar V \del \tip - V \bd \tip)
+ {\cal R} [\p_0 - \ha   \ln \L (\r)]\ \big)\ , \   $$
where $(V,\bar V)$ is a 2d gauge field and $\tip$ is a Lagrange multiplier.
Fixing the gauge $\vp=0$ and integrating out $V,\bar V$
we find the dual model   (see \mme)
\eqn\nf{{\tilde I}_3=  {1\over \pi\alpha'}\int d^2 \s\big[   \del \r \bd\r +
   \a'^2 ( \r^{-2 }  + b^2) \del \tip \bd \tip  +  \del y \bd y
+ 2\a'  b \del y \bd \tip
 + {\cal R} ({\p}_0 - \ln \r)   \big]\   }
\eqn\nff{=  {1\over \pi\alpha'}\int d^2 \s\big[   \del \r \bd\r +
   \a'^2 \r^{-2 }   \del \tip \bd \tip  +  \del \bar y \bd \bar y   +
 \a' b (\del \bar y \bd \tip - \bd \bar y \del  \tip)
 + {\cal R} ({\tilde\p}_0 - \ln \r)   \big]\ , } $$
  \ \ \ \   \bar y\equiv y + \a' b \tip \ .  $$
To  be able to argue \rocver\ that  $I_3$ in  \nnn\ and ${\tilde I}_3$ in \nf\
represent equivalent conformal
theories   we  need to impose  the condition that  $\tip$ has the same period
as
$\vp$, i.e. $2\pi$.  Then  $\bar y$ is a well-defined  periodic coordinate only
if  $\a'b/R $ is rational.
In this case
we may
 assume that  $\bar y$ is inert under   further duality rotation of
$\tip$  into $\vp'$ so that   we  end up with a  flat space model
which is a direct product of a  2-plane
and a circle\foot{A similar conclusion is reached
for the model \xxx\ by making a duality rotation
$\vp \to \tip$,  a shift of $y$ and $t$  by $\a' b \tip$
and  duality rotation  of $\tip $  ``backwards" (see \refs{\kltspl,\kk}\ where
$y$ was non-compact). Analogous  observation for the model \nnn\ with a
non-compact $y$ was made  also in  \sft, see footnote 4.
Note in this connection that the $[SL(2,R)\times
R]/R$  WZW  model
can be obtained  by an  $O(2,2)$ duality rotation  (with $q$ as a parameter)
from the  ungauged $SL(2,R)$ WZW
model \refs{\rocver,\kum}.
It may be useful also to recall
that the model of \refs{\horts,\ruts} is closely  related
 to a non-semisimple $E_2^c$ WZW model of
 \napwi\  which  itself  is a formal singular limit   \refs{\sft,\sfts}  of
 the $SU(2)\times R$ WZW model. This  suggests  that
 the models \nnn\ and \xxx\ are special cases of a more general class of
string solutions (this indeed turns out to be the case and will be discussed in
\rutsn).}
\eqn\nnf{{\tilde I}_3=  {1\over \pi\alpha'}\int d^2 \s\big[   \del \r \bd\r +
   \r^{2} \del \bar \vp \bd \bar \vp  +  \del \bar y \bd \bar y
 + {\cal R}{\tilde\p}_0    \big]\ , \ \ \ \bar \vp \equiv \vp' + by \ .   }
A shift of $\vp'$ is due to a constant antisymmetric tensor term in \nff.
The condition that $\bar \vp$ has period $2\pi$ is $bR=$integer.
This is consistent with  $\a'b/R=$rational only if
$\a'b^2$ is  also rational.

If  instead of \nnn\  one starts with the ``singular gauge" model \nn\
and makes the duality rotation of $\vp$ one  directly obtains
a flat model with constant antisymmetric tensor  and dilaton
\eqn\nfff{{\tilde I}'_3=  {1\over \pi\alpha'}\int d^2 \s\big[   \del \r \bd\r +
   \a'^2 b^2(1 + b^2 \r^{2 })  \del \tip \bd \tip
 +  \del y \bd y
- 2\a'  b \del y \bd \tip
 + {\cal R} ({\tilde\p}_0 +  \ln b)   \big]\    }
\eqn\nnff{=  {1\over \pi\alpha'}\int d^2 \s\big[   \del \r \bd\r +
   \a'^2 b^4\r^{2 }   \del \tip \bd \tip  +  \del \hat y \bd \hat y
-  \a' b (\del \hat y \bd \tip - \bd \hat y \del  \tip)
 + {\cal R} {\tilde\p}_0   \big]\ , }  $$
 \hat y\equiv y - \a' b\tip\ , \ \ \ \ \ \ {\tilde\p}_0 \equiv  {\p}_0 +    \ln
b\ .
$$
This model is equivalent to \nn\ if $\tip$ has period $2\pi$.
Then  $\hat y$ is  a globally defined compact coordinate provided
$\a'b/R =$rational.  At the same time, \nnn\ and \nn\ are equivalent if
$Rb=1/n$, i.e. if $\a'b^2$ is rational,  in agreement with what
we have found  from \nnf.  In particular, we  conclude that
 in the special case  of  $R= \a'bm/n, \ \a'b^2= 1/N$ ($m,n, N$=integers)
the Melvin CFT  \nnn\ is equivalent  to that
of an orbifold of 2-plane  $R^2/Z_N$ times  a circle.
The $R^2/Z_N$ model
 is equivalent to the  ``string on the cone" (or ``string on cosmic string
background",  or
 ``string in
 Rindler space at finite temperature")
  model  recently discussed and explicitly solved in
 \refs{\dab,\stro} (see also \refs{\bag,\sus}).  The magnetic field strength
parameter $b$ is related to  the  Rindler ``temperature"  by $\beta=
2\pi\a'b^{2}=2\pi/N, \ N=1,2, ... $.
It was found in \refs{\dab,\stro} that this model  contains tachyons (both in
the bosonic
and in the superstring versions) with masses $\a'M^2 = -4 + 4/N$.
The same
 conclusion  is thus
 true  for the Melvin model with  special value of the magnetic field strength
parameter $\a'b^2=1/N$: it
is  {\it unstable}   when considered as a
   string theory solution.\foot{Instability of the  Melvin solution (and
related absence of residual supersymmetry in the context of supergravity) was
discussed in the field-theory limit
in \gibb.}
 Similar  tachyonic instability was found  also  for the
constant  magnetic field solution \hts,\meh,\mmeh\ in \ruts.
This  suggests that uniform
magnetic field backgrounds are in general unstable in string theory.

Let us now discuss the case of generic parameters $b$ and $R$
starting with the model \nn\ or, equivalently, its dual \nfff.
The 3-metric in \nfff\  is
 \eqn\met{ ds^2=  d\r^2 +
   \a'^2 b^4 \r^{2 } d\tip ^2 + d(y-\a'b \tip)^2  =
d\r^2 + R^2 |d\psi + \tau (\r) \tip|^2 \ ,  } $$ \ \ \ y\equiv R\psi\ , \  \ \
\tau \equiv  -(1+ i b\r) \a'b/R\  .  $$
 Its constant $\r$ section is a  2-torus ($\psi$ and $\tip$ have periods
$2\pi$) with $\r$-dependent  modulus.
This  metric is regular and $flat$  for $\r\not=0$ but  degenerates
at $\r=0$ where the torus develops a pinch (Im $\tau\to 0$).  The classical
string equations which follow from \nfff\  are
\eqn\eeee{
   \del  \bd\r -
   \a'^2 b^4 \r \del \tip \bd \tip =0\ , \ \ \  \del( \r^{2 }\bd \tip)
+  \bd( \r^{2 }\del \tip)=0\ , \ \ \
  \del \bd (y- \a'  b\tip ) =0\ . }
The $\r,\tip$ equations are thus the same as in the case of the cone theory
$L= \del \r \bd\r +\a'^2b^4 \r^{2 }   \del \tip \bd \tip $.
This suggests that this theory is not well-defined for generic $b$.

The 3-metric corresponding to the model \nf\
 dual to \nnn\ is
 \eqn\mett{ ds^2=   d\r^2 +
   \a' \r^{-2 } d\tip ^2  + d(y+\a'b \tip)^2
= d\r^2 + R^2 |d\psi + \tau (\r)  \tip|^2 \ ,  } $$
\ \ \ y\equiv R\psi\ , \  \ \  \tau \equiv (b + i \r\inv )\a'/R\ .   $$
This is a curved space which is an obvious  generalisation of a 2-space dual to
2-plane. Indeed, the classical string equations for this model
\eqn\eee{
   \del  \bd\r  +
   \a'^2 \r^{-3} \del \tip \bd \tip =0\ , \ \ \  \del( \r^{-2 }\bd \tip)
+  \bd( \r^{-2 }\del \tip)=0\ , \ \ \
  \del \bd (y+ \a'  b\tip ) =0\ ,  }
are essentially the same (for $\r$ and $\tip$) as for the model
$\tilde L= \del \r \bd\r +
   \a'^2 \r^{-2 }   \del \tip \bd \tip $ and thus can be
solved explicitly in terms of free fields by observing that
duality maps classical string solutions into solutions of a  dual model
($G_{\m\n} \del_a x^\n= \ep_{ab} \del^b \tilde x_\m$, etc).
Indeed, if $\r_0$ and $\vp_0$  are solutions of the flat 2-space model
$L= \del \r \bd\r +\r^{2 }   \del \vp \bd \vp $ (expressed in terms of free
fields $x_1,x_2$ as follows  $\r^2_0= x^2_1 + x^2_2 , \  \tan \vp_0= x_1/x_2$),
then
the solution  corresponding to $\tilde L$ is
$(\r_0, \tip_0)$, where  $ \a' \del_a \tip_0  = \ep_{ab} r^2_0  \del^b \vp_0 =
\ep_{ab} (x_1  \del^b x_2 - x_2  \del^b x_1)$.
Then  also  $\del^a\del_a y = - 2  b\ep_{ab} \del^a x_1  \del^b x_2$.
Thus   \eee\ can be solved  in terms of free fields
  and  the corresponding conformal quantum theory
can be  constructed explicitly
 like it was done in  the case of \xxx\ in \ruts\ (this will discussed in
detail in \rutsn). In contrast to \nn,\nfff\  the model \nnn,\nfff\ is
well-defined
for generic $b$ and $R$ (but still  is unstable due to tachyons).

\bigskip
4. \  The direct superstring or heterotic string analogue of the  bosonic
Melvin solution is represented by $(1,1)$ or  $(0,1)$
 supersymmetric extension
of the model \nnn\  (with an internal
 gauge field background added in the heterotic  case to ensure  the global
$(1,1)$ supersymmetry, i.e. the embedding into the superstring). As in the
bosonic case, the  resulting models have the magnetic field
being of a  Kaluza-Klein origin.

One may construct a different ($D=4$)  heterotic string  version of the Melvin
solution  for  which the magnetic field belongs to  the internal gauge sector.
The idea is to  ``fermionise" the internal coordinate $y$ in \nnn.
Following
the  approach used in  \ruts\ let us
 consider   first  the $(0,1)$ super-extension of
\nnn\ in the $(\r,\vp)$ directions (adding the corresponding
 left fermions) and  then  ``fermionise" $y$  to get the  internal right Weyl
fermion coupled to
$A$ in \me\  (and an extra free left fermion).\foot{ An  attempt to get
(as in the case of  the ``monopole" theory in \gps)
 both the left and right fermions (coupled to  a spin connection of the
$(\r,\vp)$ metric and to the internal
gauge field)  from the  single field $y$
 does not work in the present case  (\nnn\ does not lead to the necessary
Thirring-type  interaction; also,   the  vierbein  connection
for the $(\r,\vp)$  part of \nnn\ $\omega^1_{\ 2} = (\L\inv-2 \L^{-2}) d\vp$
does not match  the magnetic field $A$ in \me).}
 The resulting $(0,1)$
supersymmetric heterotic \sm\foot{Here
$\psi_R $ and $\l_L =(\l_L^1 + i\l_L^2)/\sqrt2 \ $  ($\l_L^a$ correspond to the
vierbein basis $e^1= d\r, \ \ e^2 = \r\L\inv (\r)d\vp$)
are complex Weyl spinors.  We ignore extra free fermionic terms.}
\eqn\hht{
I_{(0,1)} = {  {1\ov \pi \a'} }
\int d^2 \s \big[ -  \del t \bd t
 +    \del z \bd z +   \del \r \bd \r  +  \L^{-2}  (\r)  \r^2 \del \vp \bd \vp
  -  \l_{L}^t \del \l^t_L +   \l_{L}^z\del \l^z_L  } $$
+\   {\bar \l}_{L}[\del  - i(\L\inv-2 \L^{-2})(\r) \del\vp] \l_L
 \  + \   \bar \psi_{R} [ \bd -  ie_0  b \L\inv (\r)   \r^2 \bd \vp] \psi_{R}
$$ $$
+ \ 2 e_0 b  \L\inv  (\r)  \bar \psi_{R} \psi_{R}\bar \l_L\l_L\ \big]
\   ,  \    \  \   e_0\equiv R\inv = {  \sqrt {2/\a'}}\   ,  $$
should be  conformal to all orders.
The fact that the duality transformation  in \nng\  is performed only in the
``left sector"  (in $\vp$-direction)
  suggests that  for special $b,R$ \hht\ is
equivalent to the heterotic counterpart of an  orbifold of a flat space.

The embedding of the Melvin solution into string theory we discussed above
and the existence of a similar $D=5$ representation  {\nels}
 for the extremal magnetic
dilatonic black hole background \refs{\gibma,\ghs}\foot{To
``supersymmetrically" embed this model  into the heterotic string theory one
needs also to add the internal gauge field background \kalor.}
\eqn\mag{
 I={1\over \pi\alpha '}\int d^2 \s\big[ -\del t \bd t +
F^{2}(r) \del r  \bd  r +  r^2 \del \t \bd \t
+ r^2 {\rm sin}^2 \t \del \vp \bd \vp  }
$$  +  (\del y+  M\cos \t  \del \vp)(\bd y+ M\cos \t
\bd \vp )  +  M\cos \t (\del \vp \bd y- \bd \vp \del y )  + {\cal R} \p (r)
\big] \ ,  $$
$$ \p = \p_0 + \ha \ln F(r) \ , \ \ \ \ \ F(r) \equiv
 (1 +{2M\ov r})^{-1} \ , $$
suggests that similar conformal string model
 exists for  the  dilatonic Ernst-type solution \gaun\ which
describes a pair of (extremal) magnetic black holes in a magnetic field
background and thus generalises both the dilatonic Melvin and  (extremal)
dilatonic magnetic black hole solutions.  Constructing
 such model
may open a way  to  a computation of a
  magnetic black hole pair creation rate
 \gaun\   at the level of  string theory.

\bigskip\bigskip\bigskip
I would like to thank  G. Horowitz, C.  Klim\v c\'\i k,  J. Russo
 and   K. Sfetsos for  comments and   discussions of related questions.
I  also  acknowledge the  support
of PPARC.

\vfill\eject

\listrefs
\end